\documentstyle{elsart}
\input epsf

\begin{document}

\begin{frontmatter}
\title{Degree stability of a minimum spanning tree of 
price return and volatility}

\author{Salvatore Miccich\`e$^{a,b,}$\thanksref{mail1}},
\author{Giovanni Bonanno$^{b,c}$},
\author{Fabrizio Lillo$^{a}$}, 
\author{Rosario N. Mantegna$^{a,b}$}

\address{
$^a$ Istituto Nazionale per la Fisica della Materia, Unit\`a di Palermo,
Viale delle Scienze, I-90128, Palermo, Italia\\
$^b$ Dipartimento di Fisica e Tecnologie Relative,
Universit\`a degli Studi di Palermo, 
Viale delle Scienze, I-90128, Palermo, Italia\\
$^c$ Istituto Nazionale per la Fisica della Materia, Unit\`a di
Roma1, P.le A.Moro 2, I-00185 Roma, Italia\\}

\thanks[mail1]{corresponding author, e-mail address: micciche@lagash.dft.unipa.it}

\begin{abstract}
We investigate the time series of the degree of minimum spanning trees 
obtained by using a correlation based clustering procedure which is 
starting from (i) asset return and (ii) volatility time series. 
The minimum spanning tree
is obtained at different times by computing correlation among time
series over a time window of fixed length $T$.
We find that the minimum spanning tree of asset return is characterized by
stock degree values, which are more stable in time than 
the ones obtained by analyzing a minimum spanning tree computed 
starting from volatility time series. Our analysis also shows
that the degree of stocks has a very slow dynamics 
with a time-scale of several years in both cases.


\end{abstract}


\end{frontmatter}

\section{Introduction}

The existence of correlation among price returns of different stocks traded in a financial market is a well-known fact \cite{Mark,EltGrubbook,CampbLo}. 
Correlation based clustering procedures have been pioneered 
in the economic literature \cite{EltGrub,PanPark}.
Recently, a new correlation based clustering procedure has been 
introduced in the econophysics literature.  
It has been shown that this 
correlation-based clustering procedure and some variant of it
are able to filter out information which has a direct economic 
interpretation from the
correlation coefficient matrix.
In particular, the clustering procedure is able to detect 
clusters of stocks belonging to the same or closely related 
economic sectors starting from the time series of returns only
\cite{Mantegna99,Kullmann00,Bonanno00,Bonanno01,Bonanno02}.

In this paper we will consider the problem of the
stability associated with the Minimum Spanning Tree (MST) 
obtained both from price return and volatility data. By investigating
the stability of the value of the degree (number of links of the
stock in the MST) of each stock we will show 
that volatility MST has less stable values of stock degree
than price return MST. Moreover, by analysing the
degree of elements of MSTs we will be able to show that the
degree has a slow dynamics with a correlation 
time of several years. 

The paper is organized as follows. In Sect. 2 we illustrate our results
about the MST of volatility time series of a set of stocks. In 
Sect. 3 we comment on the stability of stock degree in the MSTs of 
price return and volatility time series and we discuss the 
time-scale associated with the slow dynamics of degree of MSTs. 
In Sect. 4 we briefly draw our conclusions.   

\section{Correlation-based clustering of volatility}

We investigate the statistical properties of cross-correlation among volatility and among price return time series for the $93$ most capitalized stocks traded in US equity markets during a $12$ year time period. Our data cover the whole period ranging from January $1987$ to April $1999$ ($3116$ trading days). In the
present study we investigate daily data. In particular, we use for our analysis the open, close, high and low price recorded for each trading day 
for each considered stock. The stocks were selected by considering the capitalization recorded at August 31, $1998$. 

Starting from the daily price data, we compute both the daily price return $R_i(t)$ and the daily volatility $\sigma_i(t)$ for each stock $i=1, \cdots, 93$. Price returns are defined as
$R_i(t+1) = [P_i(t+1) - P_i(t)]/ P_i(t)$
where $P_i(t)$ is the close price of stock $i$ at day $t$. Volatility is computed by using the proxy
$\sigma_i(t) = 2~[\max\{P_i(t)\} - \min\{P_i(t)\}] / [\max\{P_i(t)\} + \min\{P_i(t)\}]$
where $\max\{P_i(t)\}$ and $\min\{P_i(t)\}$ are the highest and 
lowest price of the day, respectively. 

The correlation based clustering procedure introduced in Ref. \cite{Mantegna99} is based on the computation of the subdominant ultrametric distance  \cite{Viras} associated with a metric distance that one may obtain from the correlation coefficient. The subdominant ultrametric distance can be used to obtain a hierarchical tree and a MST. The selection of the subdominant ultrametric distance for a set of elements whose similarity measure is a metric distance is equivalent to considering the single linkage clustering procedure \cite{SLCA}. Further details about this clustering procedure can be found in \cite{MS}.

In the present investigation, we first aim to consider the MST 
associated to the correlation coefficient matrix of 
volatility time series. It should be noted that there is an
essential difference between price return and volatility probability density
functions. In fact the probability density function of price return
is an approximately symmetrical function whereas the
volatility probability density function is significantly skewed. 
Bivariate variables whose marginals are very
different from Gaussian functions can have 
linear correlation coefficients which are bounded  in a
subinterval of $[-1,1]$ \cite{Embrechts}. 
Since the empirical probability density function of volatility 
is very different
from a Gaussian 
the use of a robust 
nonparametric
correlation coefficient is more appropriate for quantifying volatility 
cross-correlation. 
In fact
the volatility MSTs obtained 
starting from a Spearman rank-order correlation coefficient are
more stable with respect to the dynamics of the degree of stocks
than the ones obtained starting from 
the linear (or Pearson's) correlation coefficient.  
The clustering procedure based on the Spearman rank-order correlation coefficient uses the volatility rank time series to evaluate the subdominant ultrametric 
distance. The time series of the rank value of volatility 
are obtained by substituting the volatility values with their ranks.
Then one evaluates the linear correlation coefficient 
between each pair of the rank time series \cite{Spear} and starting from this correlation coefficient matrix one obtains the associated MST.
\begin{figure}
\begin{center}
              \epsfxsize=6.5in
              \centerline{\epsfbox{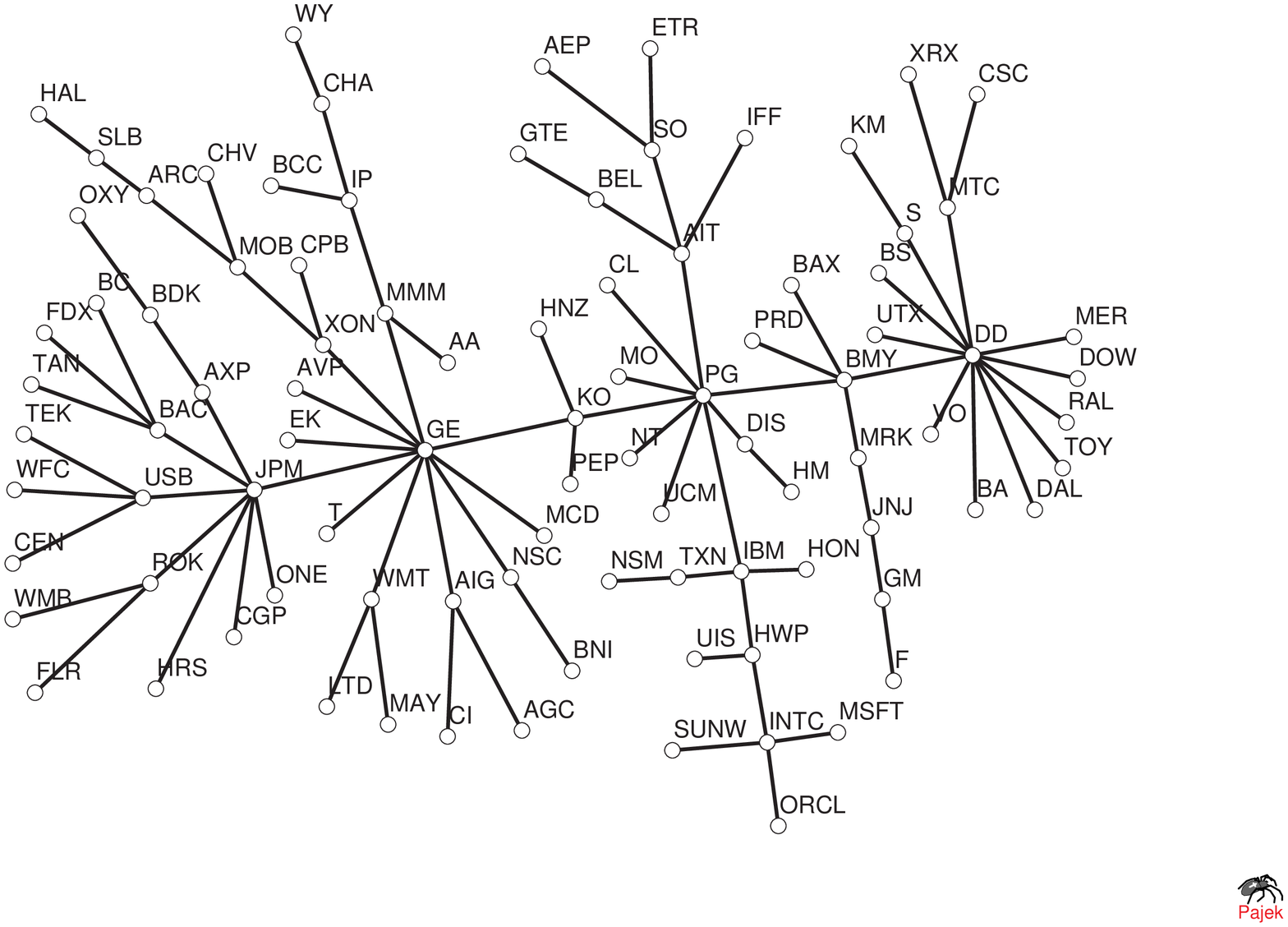}}
              \caption{Minimum spanning tree obtained by considering the volatility time series of 93 mostly capitalized stocks traded in the US equity markets in August 1998. Each stock is identified by its tick symbol. The correspondence with the company name can be found in any web site of financial information. The volatility correlation among stocks has been evaluated by using the Spearman rank-order correlation coefficient. The MST has been drawn by using the Pajek package for large network analysis http://vlado.fmf.uni-lj.si/pub/networks/pajek/ } \label{fig1}
\end{center}
\end{figure}  
An example of the MST obtained starting from the volatility time series and by using the Spearman rank-order correlation coefficient is shown in Fig. (\ref{fig1}). This MST is shown for illustrative purposes and it has been computed by using the widest window available in our database ($T=3116$ 
trading days).
A direct inspection of the MST shows the existence of well characterized
clusters. Examples are  the cluster of technology stocks (HON, HWP, IBM, INTC, MSFT, NSM, ORCL, SUNW, TXN and UIS) and the cluster of energy stocks (ARC, CHV, CPB, HAL, MOB, SLB, XON).
As already observed in the MST obtained from the
price return time series, 
the volatility MST of Fig. (\ref{fig1}) shows the existence of stocks that behave as reference stocks 
for a group of other stocks. Examples are GE (General Electric Co),
JPM (JP Morgan Chase \& Co) and DD (Du Pont De Nemours Co.).

\section{Stability of the price return and volatility MST}
A natural question arises whether or not the structure of the MST 
depends on the particular time period considered. This point has 
been considered briefly
in \cite{Bonanno00,Bonanno01} and it has also been recently
addressed in \cite{Kertesz02a,Kertesz02b}. In the present investigation
we compute a MST for both volatility and price return for each trading day $t$. This is done 
by considering the records of the time series delimited by a sliding time window of length $T$ days ranging from day $t$ to day $t+T$. For example, by using a time window with $T=117$ we approximately compute $3000$ MSTs in our sets of data. In each MST, each stock has a number of other stocks that are linked to it. This number is usually referred to as the degree of the stock. By using the above procedure, we obtain a daily historical time series of degree for each of the considered $93$ stocks both for price return and for volatility. In the following, we focus our attention on the analysis of such degree time series to assess the time stability of MSTs of price return and volatility and to infer conclusions about the time dynamics of
the stock degree of MSTs.

Each time series of the degree of each stock has about 3000 records. This
number of records is not enough to detect reliably the autocorrelation
function of the degree time series for each stock. Hence, we decide to
investigate the properties of the degree time series obtained by 
joining all the 93 degree time series of each stock. This is done
separately for each set of data (price return or volatility) and for 
each value of the time window $T$. From the time series obtained as described above we compute the autocorrelation function. The comparison of the results obtained for price return with the one obtained from volatility time series allows us to estimate the stability of MSTs of these two important financial indicators. For the sake of clarity we will first consider the two set of data separately and then we will comment on similarity and differences between them. 

\begin{figure}
\begin{center}
              \epsfxsize=5.0in
              \centerline{\epsfbox{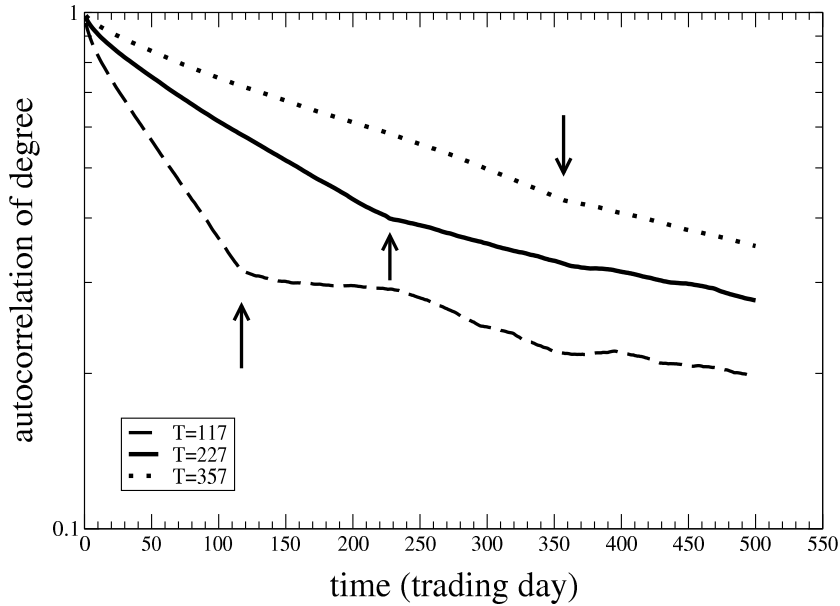}}
                            \caption{Autocorrelation function of the degree of MSTs obtained starting from price return time series by using the correlation based clustering procedure described in the text. A different line-style indicates a different value of the sliding time window. The values of the time windows are $T=117$ (dashed line), $T=227$ (solid line) and $T=357$ (dotted line) trading days. For each line, arrows indicate the length of the time window. They also indicate the point where MSTs start to be computed from non-overlapping time windows.} 
\label{fig2}
\end{center}
\end{figure}  
\subsection{Price Returns}
In Fig. (\ref{fig2}) we show autocorrelation functions of $3$ different time series of the degree. The analyzed MSTs are computed by investigating the linear correlation coefficient which is present among price return and by using three different time windows $T$. Specifically, we use time window of size $T=117$, $T=227$ and $T=357$ trading days.
In all cases the autocorrelation function shows two distinct regimes for low and high time values. The crossover between the two regimes is detected at $t=T$ (see arrows in Fig. (\ref{fig2})). For low time values the autocorrelation function of degree approximately decays exponentially (a straight line in
the semilogarithmic plot of Fig. (\ref{fig2})). This behavior reflects the fact that the sliding window used to compute MSTs contains overlapping time period of records. For this reason when a day of high correlation among several pairs of stock occurs a memory of this event remains 
within a time interval of length $T$.
This behavior is therefore simply related to the methodology used by us to compute the degree time series. 

More relevant information is obtained from the degree autocorrelation at time $t$ equal or longer than the time window size $T$.
For $t \approx T$ the autocorrelation function assumes a non negligible value approximately equals to $0.31$, $0.39$ and $0.42$ for a time window of $T=117$,  $T=227$ and $T=357$ trading days respectively. These results indicate that the information carried by the degree of the stocks in the MSTs is robust in spite of the presence of some noise dressing. The increase of the value of the autocorrelation function at $t \approx T$, which is detected by increasing $T$ indicates that
the noise dressing decreases when $T$ increases.
For time $t$ longer than $T$ the degree autocorrelation function approximately decays exponentially  with a very long time-scale $\tau_{_{R}}$. For instance, in the case of $T=117$ an exponential best fit of the autocorrelation function is obtained with the time-scale  $\tau_{_{R}}=714$ trading days. This time-scale approximately corresponds to 2.8 calendar years. It should be noticed that the $3$ autocorrelation functions obtained for different values of $T$ are approximately parallel to each other and follow an exponential function 
with the same time-scale.
Before we move to consider the analogous results obtained for volatility time series we wish to point out that the results presented in Fig. (\ref{fig2}) 
are essentially independent on the methodology used to compute the correlation coefficient matrix. In fact we obtain the same results when we use 
the Spearman rank-order correlation coefficient.

\subsection{Volatility}
\begin{figure}
\begin{center}
              \epsfxsize=5.0in
              \centerline{\epsfbox{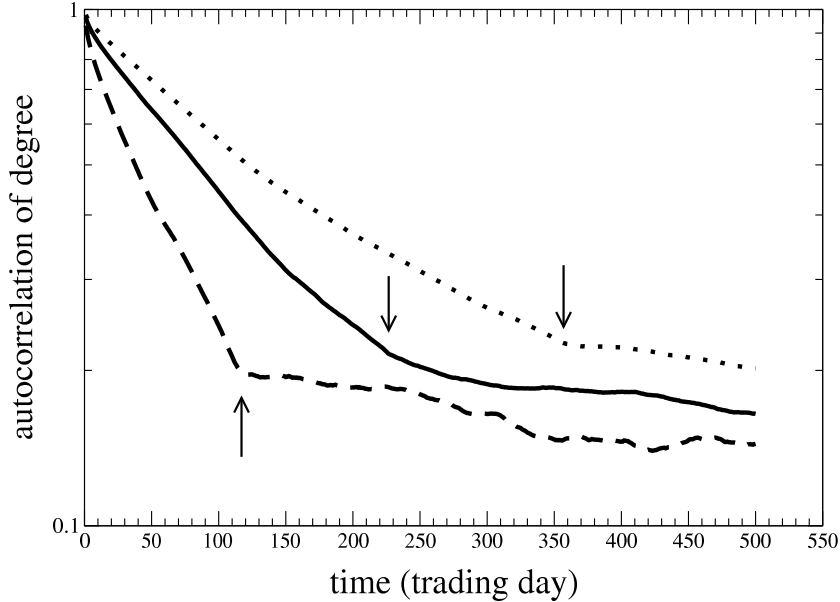}}
              \caption{Autocorrelation function of the degree of MSTs obtained starting from volatility time series by using a correlation based clustering based on the Spearman rank-order correlation coefficient and the procedure described in the text. A different line-style indicates a different value of the sliding time window. The values are the same as in Fig. (1). For each line, arrows indicate the size of the time window. They also indicate the point where MSTs start to be computed from non-overlapping time windows.} 
\label{fig3}
\end{center}
\end{figure}  
In Fig. (\ref{fig3}) we show the results of the same analysis performed on volatility time series. In the case of volatility the MSTs are obtained starting from the Spearman rank-order correlation coefficient. In fact if we compute MSTs and degree time series by using a linear correlation coefficient the results are much less reliable and the degree autocorrelation function for $t>T$ seems to be more affected by noise. MSTs obtained starting from the Spearman rank-order correlation coefficient are more statistically robust and the degree autocorrelation function shows the same general behavior as in the case of price return time series. However some important differences are detected. The first one concerns the amount of correlation observed at $t \approx T$. The values of the autocorrelation function are approximately equals to $0.19$, $0.21$ and $0.22$ for a time window of $T=117$,  $T=227$ and $T=357$ trading days respectively. These results indicate that the information carried by the degree of the stocks in the MSTs of volatility is less stable over time than the one detected in the MSTs of price returns. Moreover, the increase of the value of the degree autocorrelation time series with the time window $T$ is much less pronounced for volatility than for price return.
Another difference concerns the slow decay of the correlation observed for $t>T$. For large values of $t$ the decay of the degree autocorrelation function is again approximately exponential but the time-scale $\tau_{_{\sigma}}$ obtained by best fitting the autocorrelation function with an exponential function is $\tau_{_{\sigma}}=1510$ trading days in this time interval. This value is approximately double than the time-scale $\tau_{_{R}}$ detected in the analysis of price return.
\section{Conclusions}
The parallel investigation of MSTs obtained from (i) price return and (ii) volatility time series of a set of stocks allows us to conclude that the stability of the degree of MSTs is lower for volatility time series than for price return time series. For price return time series, the stability of stock degree dynamics increases when the time window $T$ used to compute MSTs is increased. A similar but much weaker trend is also observed in MSTs obtained starting from volatility time series. 

The dynamics of the degree of stocks in the MSTs is of  statistical nature with a time memory which is approximately close to 700 trading days for price return and 1500 trading days for volatility time series. The time-scale of the degree of MSTs of price return is much less than the maximal time-scale of our investigation $T_{max}\approx 3000$ and therefore it should not be 
significantly affected by it. On the other hand the time-scale found for the degree of MSTs obtained starting from volatility time series is just $\approx T_{max}/2$, which implies that the detection of this specific time-scale could be an artifact of the procedure we used to compute the degree autocorrelation function.
However, it should be noted that, the detected value of 
$\tau_{_{\sigma}}=1510$ trading days is certainly a lower bound of the true time-scale of the degree autocorrelation function of volatility MST.

In summary, relevant economic information is stored in the degree of MSTs obtained from price return and volatility time series. The dynamics 
of stock degree is statistically more stable for price return than for volatility MSTs and
it has a slow dynamics characterized by a time-scale of the order of 3 calendar years for price return MSTs and longer than 6 calendar years for volatility MSTs.

\section{Acknowledgments}

The authors thank INFM and MIUR for financial support. This article is part of the MIUR-FIRB project on ``Cellular self-organization nets and chaotic nonlinear dynamics to model and control complex systems". G.B. acknowledges financial support from FET open project COSIN IST-2001-33555.

\end{document}